\documentclass[10pt]{article} 
\usepackage[accepted]{tmlr}

\usepackage{amsmath,amsfonts,bm}

\def\eqref#1{equation~\ref{#1}}

\def\1{\bm{1}}

\DeclareMathAlphabet{\mathsfit}{\encodingdefault}{\sfdefault}{m}{sl}
\SetMathAlphabet{\mathsfit}{bold}{\encodingdefault}{\sfdefault}{bx}{n}

\usepackage{hyperref}
\usepackage{url}

\usepackage{lastpage}

\usepackage{url}

\usepackage[utf8]{inputenc}
\usepackage[small]{caption}
\usepackage{graphicx}
\usepackage{amsmath, amssymb}
\usepackage{amsthm}
\usepackage{booktabs}
\usepackage{multirow}
\usepackage{bm}
\usepackage{subfigure}

\usepackage{algorithm, algorithmicx,algpseudocode}
\usepackage{tabularx}
\usepackage{longtable}
\usepackage{microtype}

\usepackage{listings}					
\usepackage{color, colortbl}
\definecolor{codegreen}{rgb}{0,0.6,0}
\definecolor{codegray}{rgb}{0.5,0.5,0.5}
\definecolor{codepurple}{rgb}{0.58,0,0.82}
\definecolor{backcolour}{rgb}{0.95,0.95,0.92}
\definecolor{PythonComment}{RGB}{5,155,5} 
\definecolor{PythonString}{rgb}{0.475,0.369,0.149} 
\definecolor{PythonKeywords}{RGB}{250,75,163} 

\definecolor{codeString}{RGB}{163,21,21}

\definecolor{codeReturn}{RGB}{175,0,219}
\definecolor{codeFunctionName}{RGB}{121,94,38}
\definecolor{codeClassName}{RGB}{38,127,155}

\lstdefinelanguage{myPython}{
language = Python,
classoffset=0,
morekeywords={class, def, is, not, False, True},keywordstyle=\color{blue}\bfseries,
classoffset=1,
morekeywords={Node, str},keywordstyle=\color{codeClassName}\bfseries,
classoffset=2,
morekeywords={return, if, else, while, for, in},keywordstyle=\color{codeReturn}\bfseries,
classoffset=3,
morekeywords={__init__, make_set, union, find, find_analysis, print, append, find_with_path_compression, shatter, compress},keywordstyle=\color{codeFunctionName}\bfseries,
classoffset=4,
morekeywords={self},keywordstyle=\bfseries,
classoffset=0,
sensitive = true,
stringstyle=\color{codeString},
backgroundcolor=\color{backcolour},
commentstyle=\color{codegreen},
breakatwhitespace=false,
breakatwhitespace=false,                        
captionpos=b,                                     
numbers=none,                    
numbersep=5pt,                  
showspaces=false,                
showstringspaces=false,
showtabs=false,
basicstyle=\ttfamily\scriptsize 
}

\DeclareMathOperator{\ch}{\mathsf{ch}}

\newcommand{\ZZ}{\mathbb{Z}}

\title{N-Parties Private Structure and Parameter Learning for Sum-Product Networks}

\author{\name Xenia Heilmann$^1$ \email xenia.heilmann@uni-mainz.de\\
\name Ernst Althaus$^1$ \email ernst.althaus@uni-mainz.de\\
\name Mattia Cerrato$^1$ \email mcerrato@uni-mainz.de\\
\name Nick Johannes Peter Rassau$^1$ \email nirassau@uni-mainz.de\\
\name Mohammad Sadeq Dousti$^{2, \,}$\thanks{Work conducted during author's postdoctoral tenure at Johannes Gutenberg University Mainz.} \email msdousti@gmail.com\\
\name Stefan Kramer$^1$ \email kramerst@uni-mainz.de \\
\addr $^1$ Institute of Computer Science, Johannes Gutenberg University Mainz, Saarstraße 21, 55122 Mainz, Germany \\
$^2$ Zalando SE, Valeska-Gert-Straße 5, 10243 Berlin, Germany}

\begin{document}

\def\@fnsymbol#1{\ensuremath{\ifcase#1\or \dagger\or \ddagger\or
   \mathsection\or \mathparagraph\or \|\or **\or \dagger\dagger
   \or \ddagger\ddagger \else\@ctrerr\fi}}

\maketitle

\begin{abstract}%
A sum-product network (SPN) is a graphical model that allows several types of probabilistic inference to be performed efficiently. In this paper, we propose a privacy-preserving protocol which tackles structure generation and parameter learning of SPNs. Additionally, we provide a protocol for private inference on SPNs, subsequent to training. To preserve the privacy of the participants, we derive our protocol based on secret sharing, which guarantees privacy in the honest-but-curious setting even when at most half of the parties cooperate to disclose the data. The protocol makes use of a forest of randomly generated SPNs, which is trained and weighted privately and can then be used for private inference on data points. Our experiments indicate that preserving the privacy of all participants does not decrease log-likelihood performance on both homogeneously and heterogeneously partitioned data. We furthermore show that our protocol's performance is comparable to current state-of-the-art SPN learners in homogeneously partitioned data settings. In terms of runtime and memory usage, we demonstrate that our implementation scales well when increasing the number of parties, comparing favorably
to protocols for neural networks, when they are trained to reproduce the input-output behavior of SPNs.
\end{abstract}


\maketitle

\section{Introduction}
Privacy is the ability of individuals to be in control of who stores and processes their data. It is an ever increasing concern, and recent laws such as the EU's General Data Protection Regulation (GDPR) or the  California Consumer Privacy Act (CCPA) seek to protect the privacy of individuals. These laws strictly regulate how data about individuals should be stored, communicated, and processed.
Privacy is often at odds with machine learning (ML) techniques, where gathering a vast amount of high-quality data is the key to success. For instance, several hospitals cannot simply share their medical records and then run an ML algorithm to construct some type of diagnosis model. To overcome this, some form of privacy-preserving ML can be employed, which guarantees privacy while providing models with the same or comparable accuracy.
The problem with this approach is that commonly used protocols are often quite complex, which is due to their use of cryptographic constructs such as fully homomorphic encryption or oblivious transfer \citep{gilad2016cryptonets, al2020towards}. Furthermore, adding the cryptographic primitives often results in high computational costs \citep{gilad2016cryptonets, DBLP:journals/corr/abs-2202-02960}.

In this paper, we propose a protocol for {\em $N$-Parties Private Structure and Parameter Learning for Sum-Product Networks}. The protocol introduces an efficient scheme for structure generation, sum parameter learning and input distribution learning on sum-product networks (SPNs) with an arbitrary number of parties in a private manner. Additionally, private inference subsequent to training is possible.\footnote{Private inference with SPNs is, however, not the novel aspect of this work.}
We picked SPNs, amongst others, for the simplicity of their operations: An SPN is a graphical model that represents a probability distribution. It is a rooted directed acyclic graph (DAG), consisting of two types of inner nodes: sum and product nodes. The sum nodes compute a weighted sum over the values of their children, where the weights are trainable and defined by the edges connecting a sum node to its children. The product nodes simply multiply the values of their children. SPNs have been described as a middle ground between Neural Networks and Probabilistic Graphical Models such as Bayesian Networks, with a performance that has been approaching that of Deep Neural Networks (DNNs) \citep{RAT-SPN}.
To achieve efficient parameter learning for SPNs in a privacy-preserving setting, we take advantage of {\em secret sharing} approaches to secure multi-party computation instead of homomorphic encryption.
\begin{figure}[t]
    \centering
    \includegraphics[width=0.65\columnwidth]{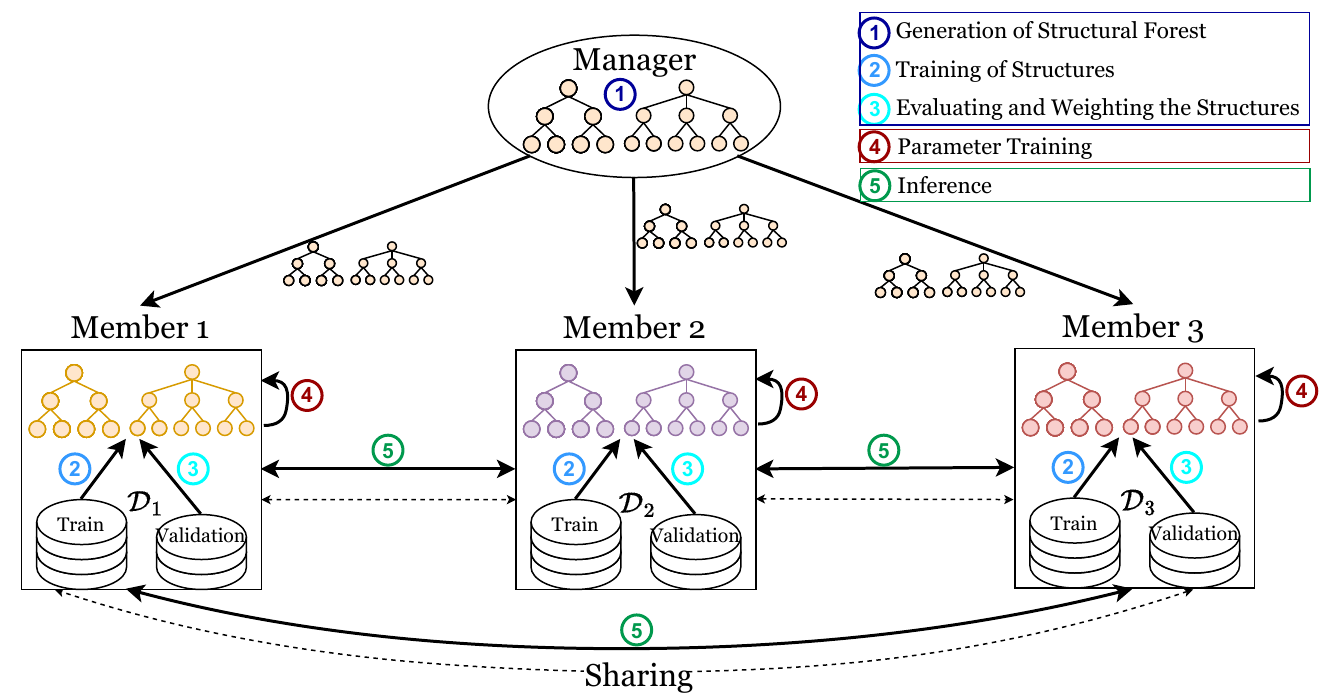} 
    \caption{Overview of the whole privacy-preserving training and inference protocol: 1. Generation of the SPN forest. 2. Training of the structures on the local training sets. 3. Evaluation and weighting of the structures on the local validation sets. 4. Private sum parameter and leaf parameter training. 5. Private inference of local data points.}
    \label{fig:overview}
\end{figure}
The idea for our protocol for N-parties private structure and parameter learning for sum-product networks is, in short, to use a forest of several randomly generated SPN structures \citep{RAT-SPN}, evaluate and weight these locally and then train shared parameters on all structures. 

To be more precise, we address the following problem: For a dataset $\mathcal{D}$, which is horizontally distributed across $N$ clients, a protocol for structure and parameter learning of SPNs is constructed. Throughout this protocol, no private data is revealed or shared among the clients and no party $n$ learns anything about the data $\mathcal{D}_j$ of party $j$ for all $n\neq j$. Our solution makes use of a weighted forest of SPN structures, which are generated without the need of a joint training set, but are randomly initialized and weighted according to their performance. Additionally, all parameters are trained privately for all structures in the forest. At the end of the protocol, each participant holds a secret share of each parameter (structure weights, sum parameters and leaf parameters). These parameters can then be used for inference on private data points. A graphical representation of our protocol can be found in Figure~\ref{fig:overview}. 

There are many protocols for privacy-preserving parameter learning based on similar secret sharing methods for different models, such as DNNs \citep{TanKTW21, Piranha} or clustering methods \citep{HegdeMSY21}. A main advantage of our protocol compared to most others is that it is not specialized to two or three parties. Hence, it guarantees privacy even if at most half of the parties cooperate to disclose the data. 


\section{Related Work}	\label{sec:prelim}
\subsection{Sum-Product Networks (SPNs)}

A sum-product network (SPN) \citep{poon2011sum} is a rooted acyclic graph representing a multivariate probability distribution. It consists of leaf nodes (representing probability distributions over random variables) and inner nodes performing sum or product operations on their children. SPNs may be employed on both discrete and continuous data. As stated above, all leaf nodes are distribution nodes with distribution functions over a subset of random variables. These can be either a PMF (discrete random variables ), a PDF (continuous random variables), or also mixed distribution functions. This paper focuses on the first case.

Sum nodes calculate a weighted sum of their children’s values, where the outgoing edge weights sum to 1. Product nodes multiply the values of their children. These layers alternate until reaching a root sum node, whose value represents the network’s output.  The scope of a node is defined as the set of variables it depends on: for leaf nodes, this is the variable set of its distribution; for inner nodes, it's the union of the scopes of its children.

SPNs may be interpreted as a subclass of probabilistic graphical models (PGMs) in which inference is tractable due to the graph's structure.
An SPN may be employed to perform probabilistic inference in time linear in the number of nodes of the network \citep{poon2011sum}.
Additionally, several other types of inference can be performed in time that is a polynomial function of the number of edges in the graph \citep{poon2011sum, sanchez2021sum}. 
These efficient types of inference include computing marginal and posterior probabilities, most probable explanation (MPE), approximate maximum a-posteriori (MAP), and approximate MAX. For a detailed introduction to SPNs and these types of inference, the reader is referred to a survey paper by \cite{sanchez2021sum}.

\subsection{Private SPNs}

Despite their impressive performance \cite{RAT-SPN}, only few privacy-preserving protocols for SPNs exist. To the best of our knowledge, there is only one previous work regarding privacy-preserving {\em inference} on SPNs, called \textsc{CryptoSPN} by \cite{cryptospn}. It uses Yao's garbled circuit \citep{yao1982protocols} for a two-party inference protocol between a client and a server. Here, the client queries the server which holds a trained SPN model, thus receiving the result of the inferred data. CryptoSPN ensures that neither party learns the other's private information. While efficient, CryptoSPN's two-party limitation restricts its practical applicability. Our work advances the field by enabling joint, private learning of both the SPN model structure and parameters in an N-party setting, followed by private inference.

Another work in the field of privacy-preserving SPNs is \textsc{UnlearnSPN} by \cite{becker2022certified}. In this work, the authors design an algorithm which removes the influence of single data points from a trained SPN to comply with legal regulations such as \textit{the right to be forgotten} (GDPR). Their experiments show that this method proves superior to retraining a full SPN each time a data point is removed from the initial dataset.

Lastly, DPSPNs \citep{heilmann2024differentially} train a differentially private SPN which can then be applied for differentially private inference, to generate differentially private synthetic data or perform marginal inference over the class labels. Differently to our method, DPSPNs rely on the privacy concept of differential privacy \citep{DworkDP}. 
\section{Secure Multi-party Computation}\label{sec:ss}

Data for learning accurate ML models is often split or distributed across different organizations or institutions, and access to it is restricted by regulations (such as GDPR) or competition between companies. To still take advantage of this distributed and often high-quality data, a solution is secure multi-party computation (SMPC).
SMPC, in short, is a cryptographic technique that enables multiple parties to compute a function on their inputs, without revealing the inputs to each other. 


\subsection{Properties}
SMPC enables $N$ parties with private inputs $x_1,\ldots,x_N$, respectively, to jointly compute a function $f$, such that $(y_1,\ldots,y_N) = f(x_1,\ldots,x_N)$. 
Informally, this computation is called secure if at the end of the protocol, each party $n$ learns nothing beyond its own private input and output $(x_n, y_n)$. More formally, $f$ is secure if for each party $n$ there exists a probabilistic polynomial-time machine $Sim$ that, on input $(x_n, y_n)$, \emph{simulates} the view%
\footnote{The \emph{view} is the set of messages received by a party during the protocol.}
of party $n$ in the protocol. In other words, each party could have computed messages with the same or at least an (computational) indistinguishable\footnote{Two probability distributions are called (computationally) indistinguishable, iff for any probabilistic (polynomial time) Turing machine, the probability of an output $1$ is the same (up to a negligible function) for both distributions.} probability distribution over messages, given it knew its private input and output, as it receives in the protocol. Hence, no knowledge is gained by the communication during the protocol (a formal definition can be found in \cite{goldreich2004foundations}, Section~7.2.1). As such, preserving privacy is one of the most important properties of SMPCs.

SMPC protocols are often designed assuming parties are \emph{honest-but-curious} (following the protocol but attempting to gain knowledge into other parties' private inputs), and can be extended to a \emph{malicious} setting (parties may arbitrarily deviate from the protocol) via verification at each step \citep{goldreich1987play,chor1985verifiable}. This is why our protocol assumes that all parties are honest-but-curious.

In this paper, we use secret sharing for SMPC: each party splits its private input into several shares, and distributes these among other parties (see Section~\ref{sec:polyss}). This setup guarantees that a share does not reveal any information about the private input of that party. However, by putting together all shares, it is still possible to reconstruct the original private input.%
Throughout the protocol, parties compute locally on their shares and exchange shares of intermediate results. That is, at no stage are the intermediate results themselves revealed. Finally, shares are combined to reconstruct the final output.


\subsection{Polynomial Secret Sharing}\label{sec:polyss}
\begin{figure}[t]
    \centering
    \includegraphics[width=0.8\columnwidth]{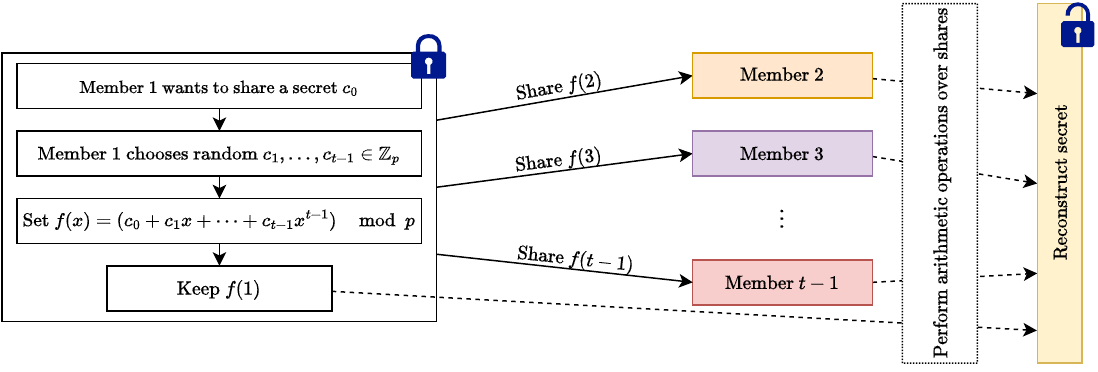} 
    \caption{Visualization of the $t$-out-of-$n$ polynomial secret sharing scheme for secret $c_0$ and with a given prime $p$.}
    \label{fig:secret-sh}
\end{figure}

In our protocol we use polynomial shares over $\ZZ_p$ \citep{shamir1979share} for secret-sharing with $p$ a prime number. Here, for an $t$-out-of-$n$ secret sharing scheme, a random polynomial of degree $t-1$ is constructed: $f(x) = c_0 + c_1x + \cdots + c_{t-1}x^{t-1} \pmod p$, where $c_0$ is the secret, and $c_1,\ldots, c_{t-1}$ are picked uniformly at random from $\ZZ_p$. The share of party $n$, $1 \le n\le N$ equals $f(n)$. With the shares of $t$ parties, the polynomial and hence $f(0)$ can be reconstructed, but with a subset of the shares, $f(0)$ remains hidden. This is the case, as $t$ points on a polynomial of degree $t-1$ completely determine it. Mathematically speaking, this can be done using the Lagrange interpolation formula. A visualization of how polynomial secret shares are created and distributed can be found in Figure~\ref{fig:secret-sh}.

In the following, all computations are done within the field $\ZZ_p$. $\lfloor d/b \rfloor$ denotes the residue of the number obtained when computing within the rationals. We use this for computations in the integers by guaranteeing that all numbers computed are in $\{0, 1, \dots, M\}$ for a number $M<p$ and hence the residue uniquely determines the number we want to compute. We shortly review basic protocols for polynomial sharing and sketch two changes we made to the protocol proposed by \citep{algesheimer2002efficient}. Firstly, the division by a shared number $x$ is done by computing a scaled inverse of $x$, i.e.,~$\lfloor d/x \rfloor$ for a suitable choice of a public number $d$. The protocol of  \cite{algesheimer2002efficient} assumes that a number $u$ such that $u/2 \le \lfloor d/x \rfloor \le u$ is known. We show that $\log M$ additional iterations of the protocol suffice, if we only know that $1 \le x \le M$. Secondly, the protocol for the division uses another protocol for an approximate division by a public number, i.e.,~given a shared number $x$ and a public number $d$, we compute a number $y$ with $\lfloor x/d \rfloor \le y \le \lfloor x/d \rfloor + N$. We simplify the protocol given by \cite{algesheimer2002efficient}, while preserving the same security guarantees.
Although developing new basic protocols was not the focus of our research, we think that these modifications can be of general interest.  
For additions and multiplications we apply existing basic protocols (see Appendix \ref{app:SMPCadd+multi}).

\subsubsection{Custom Division} 
Here, we shortly introduce our custom modifications to the division protocol proposed by \citep{algesheimer2002efficient}. 
For an exact division over shares, we compute an approximate inverse of a shared number scaled with a public normalization factor 
(adaptation of a method by \cite{algesheimer2002efficient}). In the following, let us denote $d$ as our public normalization factor. 
Note here that $d$ adjusts the precision of the division. Given this protocol, the parties owning shares of $a$ can perform a secure multiplication to compute shares of $a \cdot \lfloor d/b \rfloor$. Later, when they want shares of $\lfloor a / b \rfloor$, they can, for example, simply perform a secure truncate to remove $d$. Secure truncation refers to a protocol which on input of a shared number $b$ and a public number $d$ outputs polynomial shares of $y$ such that $\lvert y-\lfloor b/d \rfloor \rvert \leq N + 1$.

We now give a simplified version of the protocol for this secure truncate compared to the one in \cite{algesheimer2002efficient}.
The protocol of \cite{algesheimer2002efficient} is quite involved and later used to compute an approximate modulo, i.e., a number $y$ with $y \mod d=b \mod d$ and $0 \le y \le N\cdot d$. We give a protocol to compute such an approximate modulo in the next paragraph. Then, we compute the approximate value $\lfloor b/d \rfloor$ by $(b-y) \cdot d^{-1}$ within the finite field. Here, we use that $(b-y)/d$ in the rational numbers and the smallest non-negative representative of $(b-y)\cdot d^{-1}$ in the finite field $\ZZ_p$ are the same.
To compute the approximate modulo, each party $i$ generates a random number $r_i \in_R \{0, 1, \dots, \lfloor (p-M)/N \rfloor -1 \}$ and distributes shares of $r_i$ and of $m_i=r_i \mod d$. The sum $S=b + \sum r_i$ is computed and revealed to one party. This party computes $S'=S \mod d$ and distributes shares of the result. Now $\sum m_i - S'$ is an approximate modulo of $b$.

For describing the private division protocol, we denote a number $x$ which is only known in polynomial shares over $\ZZ_p$ with $[x]_p$. As in \cite{algesheimer2002efficient}, we use Newton's method (in rational numbers) to compute an approximate root of the function $f$ with $f(x)=1/x - b/d$ which is an approximation of $[u]_p \approx d/b$. Recall that a single step of Newton's method given a current approximation $u$ is $u \leftarrow u-f(x)/f'(x)=u+(1/u-b/d)/(1/u^2)=u(2-ub/d)$. Notice that the division causes a rounding error, if we stick to integers.
If we start with an approximate $u$ with $u/2 \le d/b \le u$ and increase the precision in each round by a certain factor $2^t$, we have a quadratic convergence \citep{algesheimer2002efficient}. More precisely, they replace an iteration by $u \leftarrow u(2-ub/(d2^t))$ and showed that for any $t > \log(5+\ln(k+1))$, after at most $\log(t)$ steps, we have an approximation of $d/b$ with a relative error of at most $16(k+1)/2^t$, if the difference of the computed division by a public number $d$ and the real number is at most $k+1$.

As we do not have a number $u$ with $d/2b \le u \le d/b$, we start with a small $u$; i.e.,~an underestimation of $u$. After $\lceil \log(d) \rceil$ iterations, we have $d/2b \le u \le d/b$. This can be seen as follows. Let $u_i$ be the approximation of $d/b$ after the $i$-th iteration and $f_i=d/bu_i$. As $u_{i+1}=u_i(2-u_ib/d)=u_i(2-1/f_i)$, we have $f_{i+1}=f_i f_i/(2f_i-1)$. Hence, if $f_i \le 2^n+1$, we have $f_{i+1} \le 2^{n-1}+1$. Since $f_0 \le d+1$, we have $f_{\lceil \log(d) \rceil} \le 2$.
Therefore, we perform Newton's method for additionally $\lceil \log(d) \rceil$ steps. This holds even with the truncation method that only approximates the division by $d$, as we can easily modify our protocol to only overestimate the result by at most $N$.

 \section{ Privacy-Preserving Structure and Parameter Learning} \label{sec:wholeprotocol}
Our protocol for privacy-preserving structure and parameter learning for SPNs is visualized in Figure~\ref{fig:overview} and its pseudo-code is provided in Algorithm~\ref{alg:cap}. It is generally divided into three steps: Structure learning, Parameter training, and Inference.

For structure learning, we firstly generate the structural forest (step~1 in Figure~\ref{fig:overview}) and send it to each client. Then, all clients partition their local data into a training set and a validation set and train the structures (step~2 in Figure~\ref{fig:overview}) on their training data. Subsequently, the clients evaluate the structures' performance on their local validation set and weight the structures according to this performance (step~3 in Figure~\ref{fig:overview}). These weights are then distributed as shares over all clients so that each client holds a share of a global weighting vector for the structures.
For parameter training, the input distributions and sum parameters of the structures are trained privately together (step~4 in Figure~\ref{fig:overview}). After this step, each client holds private shares of all parameters and private inference on local data points is possible. For inference, each data point is evaluated on all structures and the result is weighted by the previously defined weighting vector for the structures (step~5 in Figure~\ref{fig:overview}).
\begin{algorithm}[t]
\footnotesize
\caption{Private SPN Training}\label{alg:cap}

\begin{algorithmic}[1]
\Require{$\mathcal{F}$: dictionary to generate the SPN forest; $network$: a network with $N$ members; $\mathcal{I}_{sec}$: parameters for polynomial sharing; $i$: iteration steps for local training.}
\Ensure {the secret shares for the structure weights, sum parameters and leaf parameters. }
\State setup\_network\_communication($network,\, \mathcal{I}_{sec}$)
\State F = generate\_SPN\_forest($\mathcal{F}$) \Comment{Section~\ref{sec:ratspn}}
  \For{$n \in [1, \dots, N]$}
    \State forest\_loglikl =[ ]
    \For{$\mathcal{S}\in$\,F}
        \State local\_EM\_optimization($\mathcal{S}$, $n$.training\_data, $i$)
        \State forest\_loglikl.append(mean(log\_likelihood($\mathcal{S}$, $n$.evaluation\_data)))
    \EndFor
    \State calculate\_structure\_weights(forest\_loglikl, $n$) \Comment{Section~\ref{sec:forestweight}}
    \State share\_structure\_weights(forest\_loglikl, $n$) \Comment{Section~\ref{sec:polyss}}
  \EndFor
  \For {$\mathcal{S}\in$\,F}
    \State compute\_sum\_parameters($\mathcal{S}$) \Comment{Section~\ref{sec:sumparams}}
    \State compute\_leaf\_parameters($\mathcal{S}$) \Comment{Section~\ref{sec:leafparams}}
    \EndFor
\end{algorithmic}
\end{algorithm}

\subsection{Privacy-Preserving Structure Learning}

For privacy-preserving SPNs, structures built without a joint training dataset are crucial, as data is often distributed and subject to privacy regulations. Random and tensorized sum-product networks (RAT-SPNs) \citep{RAT-SPN} were an early approach to generating such structures, often achieving comparable or superior log-likelihoods to more complex learners like \textsc{ID-SPN} \citep{ID-SPN} or \textsc{LearnSPN} \citep{LearnSPN}. We leverage RAT-SPNs, but employ a weighted forest of them instead of a single structure. This mitigates the risk of poor performance from a randomly generated RAT-SPN and improves overall robustness and accuracy.

\subsubsection{Structures in the Forest} \label{sec:ratspn}
RAT-SPNs are constructed in two steps. The first step is the building of a region graph $\mathcal{R}$ \citep{regiongraph1, regiongraph2}, which then serves as the basis for a corresponding SPN. A region graph is an abstract representation of a network structure constructed by randomly splitting the root region which contains the whole set of random variables $\mathbf{X}$ into two sub-regions of equal size. This procedure is then repeated until a predefined split depth $D$ is reached. This mechanism as a whole can then be applied a fixed number of times, which is later denoted as the number of recursive splits $R$. 
From the region graph $\mathcal{R}$, the SPN can be constructed by populating the root region with $\mathcal{C}$ sum nodes (here $\mathcal{C}=1$ is fixed). Then, for all regions which represent sum nodes, a fixed number $S$ of sum nodes is created. On the leaf level, each leaf region is populated with a fixed number $I$ of input distributions. 

In our protocol, we furthermore transform the RAT-SPNs such that they fulfill the two structural properties: (1) \textit{Completeness:} All sum nodes have children whose scopes are the same (i.e., defined over the same model variables), and (2) \textit{Decomposability:} All product nodes have children whose scopes are disjoint. Additionally, our transformed RAT-SPNs have the property of (3)~\textit{Selectivity} \citep{peharz2014learning}: For each sum node, at most one child makes a positive contribution. 

Before local training, we generate a forest of RAT-SPN structures and distribute it to all clients (also see Figure \ref{fig:overview}). Concretely, we have that after this step all clients hold identical structures with the same of number of nodes, layer depth and node types. The clients then optimize the received structures' parameters based on their local dataset for the subsequent step of weighting the forest. Importantly during this optimization as well as the later parameter learning, the forests' structures always stay the same terms of number of nodes, layer depth and node types, only their parameters and weights are updated. 
 \subsubsection{Weighting the Forest} \label{sec:forestweight}

The contributions of the SPNs in the forest can be combined in a way that each SPN is given equal weight. In the following, we call this {\em uniform weighting}. As the structures are generated randomly, it is to be expected that some of the SPNs perform poorly, which then also affects the performance of the uniform weighting scheme. Therefore, it makes sense to weight the SPNs by their performance, i.e., by their log-likelihood. A further step is to weight the SPNs not by their (normalized) log-likelihood, but by their relative performance. This is useful as the log-likelihoods may vary largely. It is the simplest scheme that takes into account the differences in log-likelihood, however, not according to their actual quantity, but according to their rank. The uniform weighting and the log-likehood based weighting schemes are described and evaluated in full detail in Appendix \ref{apx:impl}. 


In the following, we elaborate on the log-likelihood based weighting scheme that is based on ranks. 
For weighting the generated structures there are $N$ parties which each hold a local dataset $\mathcal{D}_n$, which is split into a dataset for training and a dataset for validation. Furthermore, all hold the same structural forest with $K$ structures $\mathcal{S}_1, \dots ,\mathcal{S}_K$. These structures are then trained by each client $n$ on their local training set with the expectation maximization (EM) algorithm \citep{augmentation}. After training, since each client holds a different local training set, the structures have varying sets of trained parameters for each client and are therefore denoted as $\mathcal{S}_1^n, \dots ,\mathcal{S}_K^n$ in the following. All structures are then evaluated in terms of log-likelihood on each client's local validation set and are given a rank ${r}_k^n$ from $1$ to $K+1$. Here, higher ranks indicate a better performance on the validation set. Once ${r}_k^n$ is defined for all structures, the local structure weight $s_k^n$ can be calculated by
\begin{equation}
    s_k^n = \frac{r_k^n}{\sum_{k=1}^K r_k^n}.
\end{equation}
It holds that $\sum_{k=1}^K  s_k^n = 1 $ and higher ranks result in greater structure weights. In the next step, each party $n$ sends shares $ \widehat{s}_k^n$ of all locally computed weights divided by the number of clients, so shares of $\frac{1}{N}{s}_k^n$, to all other clients. Having received all shares $ \widehat{s}_k^n$ from all clients for each structure, each client locally takes the sum  $\sum_{n=1}^n \widehat{s}_k^n$ for each structure $k$. The result of this sum is then party $n$'s share of the weight $s_k$ for structure $k$. Because of the earlier division by $\frac{1}{N}$, it holds that $\sum_{k=1}^K s_k=1$.

Notice that in the protocol most parameters and values are in $[0,1]\in\mathbb{R}$. To map them into integers, which is needed for secret sharing, we choose a normalization factor $d$ and multiply all numbers with it. Therefore, we assume throughout this section that all numbers that are given into our protocol are integers in $[0, \dots, d]$, mostly given in polynomial shares. We guarantee that the results are again integer numbers in $[0, \dots, d]$. We assume that the prime $p$ is greater than $d$.

\subsection{Privacy-Preserving Parameter Learning} \label{sec:params}
After the initial local training process and the weighting of the forest, the next step is to calculate shared sum and leaf parameters for each structure in the forest. This is done as described in the following.

\subsubsection{Sum Parameters} \label{sec:sumparams}
Let us denote the nodes of an SPN by $m_1, m_2, \ldots$. Whenever there is an edge $m_i \to m_j$, we call $m_j$ the child of $m_i$. The set of children of $m_i$ is denoted by $\ch(m_i)$. If $m_i$ is a sum node, the outgoing edge to any of its children $m_j$ has a weight $w_{ij}$. We assume the weights are non-zero, and the sum of all outgoing weights from a sum node is 1.  

For a structure $\mathcal{S}_k$ in the forest, its sum nodes $m_{i,k}$ and any $m_{j,k}\in \ch(m_{i,k})$, define $m_{ij,k}$ as the number of instances for which the child node $m_{j,k}$ makes a positive contribution to its parent $m_{i,k}$. Similarly, for the structure $\mathcal{S}_k^n$ which is trained locally by party $n$ the variable $m_{ij,k}^n$ defines the number of instances for which $m_{j,k}^n$ makes a positive contribution to its parent $m_{i,k}^n$. Then the weight  $w_{ij,k}$ on the edge from $m_{i,k}$ to $m_{j,k}$ is defined as:
\begin{equation}\label{sumw}
    \centering
    w_{ij,k} = \frac{ \sum_{n=1}^N m_{ij,k}^n}{ \sum_{n=1}^N den_{ij',k}^n}.
\end{equation}
Here, the numerator is the sum over all clients and their locally computed number of instances for which a fixed child $m_{j,k}^n$ locally makes a positive contribution to $m_{i,k}^n$. The denominator is the sum over all numbers of instances for which an arbitrary child $m_{j',k}^n$ makes a positive contribution to the sum node $m_{i,k}^n$ locally on a party $n$. Concretely, it holds that 
\begin{equation*}
    den_{ij',k}^n = \sum_{{m}_{j',k}^n \in \,\ch(m_{i,k}^n)} {m}_{ij',k}^n.
\end{equation*} The division in Equation~\ref{sumw} is performed in a private manner as described in Section~\ref{sec:polyss}. After the private division protocol, all clients then hold a share of each weight $w_{ij,k}$ locally on their machine.

\subsubsection{ Leaf Parameters}\label{sec:leafparams}

This paper focuses on discrete probability distributions over a random variable $X$ with binary values $\text{val}(X) = \{0,1\}$. Concretely, the leaves in the structures $\mathcal{S}_1, \dots ,\mathcal{S}_K$ represent Bernoulli distributions, with a trainable parameter $\mathit{leaf.p}$. It holds that $\mathrm{P}(X =1) = \mathit{leaf.p}$ and $\mathrm{P}(X =0) = 1-\mathit{leaf.p} = \mathit{leaf.q}$.
After local training, each party $n$ holds the structures $\mathcal{S}_1^n, \dots ,\mathcal{S}_K^n$ with optimized input distributions. Now, let $\mathit{leaf.p}^n_k$ represent one trained parameter of the input distribution for the local SPN $\mathcal{S}_k^n$. Then, to calculate a global parameter $\mathit{leaf.p}_k$ for each leaf node, the mean over the trained parameters of all $N$ parties is taken: 
\begin{equation}
    \mathit{leaf.p}_k = \frac{1}{N} \sum_{n=1}^N \mathit{leaf.p}^n_k.
\end{equation}
This can be done in a private manner, similar to the structure weight calculation in Section~\ref{sec:forestweight}. 
To ensure that private inference over shares is possible subsequent to training, the same procedure as described for the parameter $\mathit{leaf.p}_k$ is also performed for $\mathit{leaf.q}_k$ for all leaves. As a consequence, after the training of the input distribution is finished, each client holds a share of $\mathit{leaf.p}_k$, and $\mathit{leaf.q}_k$ locally on its machine.

Sharing the leaf parameters has the advantages that external clients, which have not taken part in the learning process and therefore hold no SPNs locally on their machine, can take part during inference. These clients would only have to distribute shares of their data point to all clients holding shares of the parameters and could thus conduct inference for this data point. 
\begin{figure}[t]
    \centering
    \includegraphics[width=0.6\columnwidth]{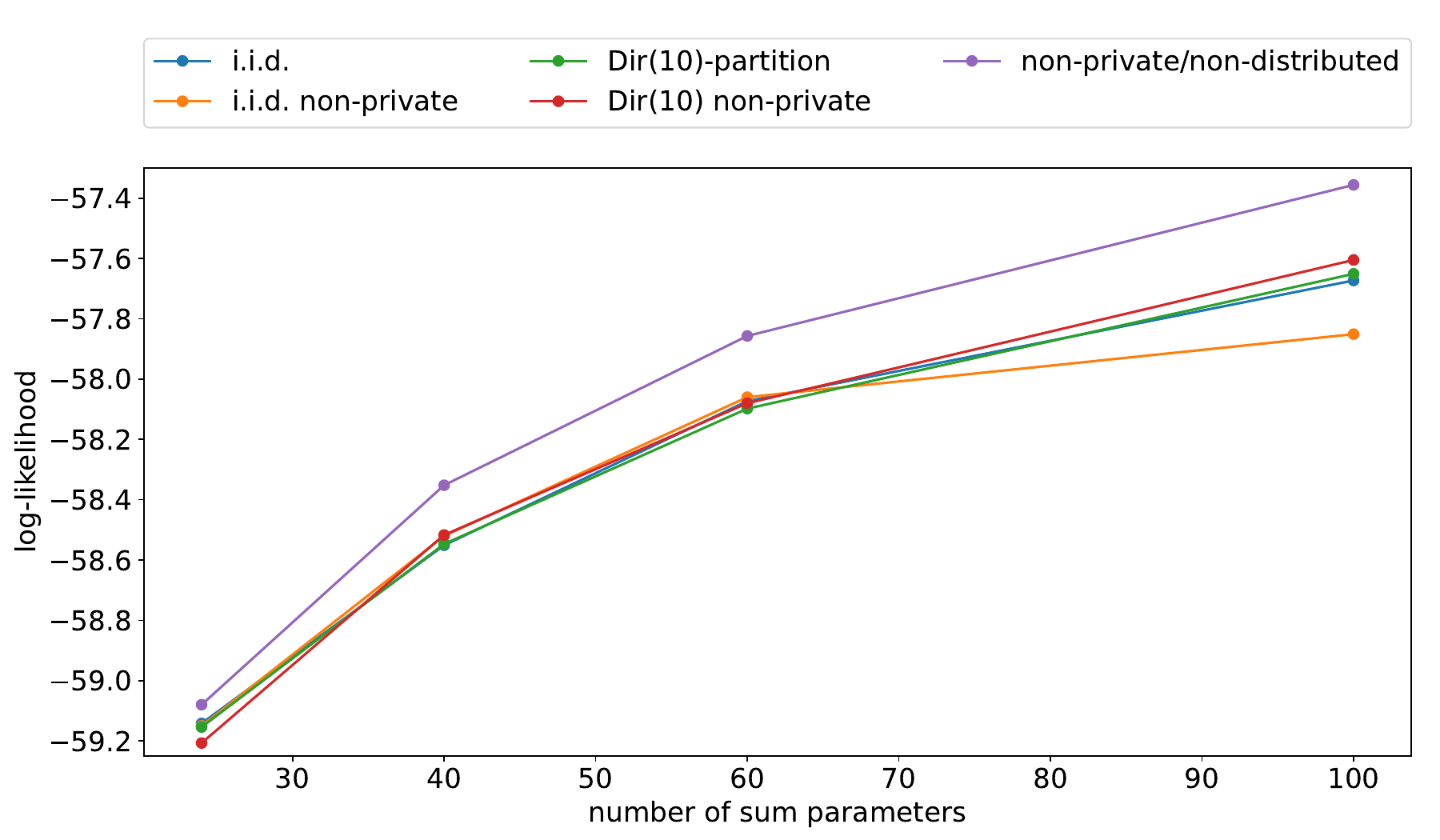} 
    \caption{\textbf{Comparison} of the different non-private baselines with our privacy-preserving protocol. Mean log-likelihood performance of 10 runs on the bnetflix dataset with 3 members and a varying sum parameters on i.i.d. data and $Dir(10)$ partitioned data. The log-likelihoods of the four distributed partitions are all very close, only the non-private, non-distributed setting performs slightly better. Full tables can be found in the Appendix in Tables~\ref{tab:all-lls-iid}, \ref{tab:all-lls} and \ref{tab:all-lls-np}.}
    \label{fig:ll-np}
\end{figure}

\subsection{Privacy-Preserving Inference} \label{sec:infer}
Our protocol also enables that inference can be done privately once the training process has finished, i.e., once all parameters are distributed as shares on the clients. During private inference no previously learned share of the parameters should be revealed to other clients. Additionally, no party should learn anything about the input data if it is not their own. We here focus on computing the probabilities $P(\mathbf{x})$ of input states $\mathbf{x}$. These are calculated in the root value $\mathbf{S}(\cdot)$ of an SPN $\mathcal{S}$ by an upward pass from leaf nodes to the root node in time linear to the number of links in $\mathcal{S}$. However, other types of inference can be realized in the future with only minor changes to the current implementation.    

If party $n$ wants to perform a private evaluation on a binary input $\mathbf{x}$, the client first has to distribute shares of the input to each other client. Now each client calculates with its share of the input distributions a share of the leaf value. Once this calculation is done for all leaves in a structure, the value of the sum nodes and product nodes can be determined by secure summation and secure multiplication (see Section~\ref{sec:polyss}). For the product nodes, a secure multiplication over the shares of their children's values is performed. The value of the sum nodes is calculated by a secure summation over the shares that lie locally on the clients after they performed a secure multiplication over the shares of all sum nodes' children and their corresponding weights. 
After this procedure is finished for all structures $\mathcal{S}_1, \dots ,\mathcal{S}_K$, a final secure multiplication over the shares of the calculated root values $\mathbf{S}_k^n(\mathbf{x})$ on the clients and their shares of the structure weights is performed. This results in a final share $\mathbf{S}^n(\mathbf{x})$ on each party $n$ of the global result $\mathbf{S}(\mathbf{x})$. Share $\mathbf{S}^n(\mathbf{x})$ is then sent to the client who initiated the inference round and after putting all shares together, thus receives the clear value of $\mathbf{S}(\mathbf{x})$. 



\section{Experiments}	\label{sec:exp}
The goal of the experiments is to perform density estimation while measuring the log-likelihood performance, the required time and network traffic for private training of all parameters in the forest and the scalability as the number of parties increases. We additionally evaluate the log-likelihood performance of our protocol on heterogeneously partitioned data and provide a training time comparison with an SMPC trained neural network model as well as two non-private versions of our protocol. Furthermore, we show the expected time for inference with regard to different numbers of participating clients. 
We implemented our protocol with Python and the code may be found at 
\url{https://github.com/xheilmann/ppsl-spn/tree/main}.
Implementation details and hyperparameter settings can be found in Appendix \ref{apx:impl}. 

We evaluated our approach on five datasets: nltcs, plants, baudio, bnetflix, and tretail \citep{datasets, tretail}. We tested both homogeneous (i.i.d.) and heterogeneous data distributions across clients. For heterogeneous distributions, we used a $K$-means clustering approach to create data partitions and simulate varying levels of imbalance using a Dirichlet distribution $Dir(\beta)$, with $\beta$ values of $0.5$ (strong imbalance) and $10$ (weak imbalance). We also simulated a very strong imbalance setting by assigning to each client only data from a single cluster. 
All experiments were repeated 10 times using a forest of 3-layered structures with varying numbers of sum weights. We also evaluated eight different local training criteria on the smallest forest (details in Appendix~\ref{apx:impl}).

\begin{figure}[t]
    \centering
    \includegraphics[width=0.6\columnwidth]{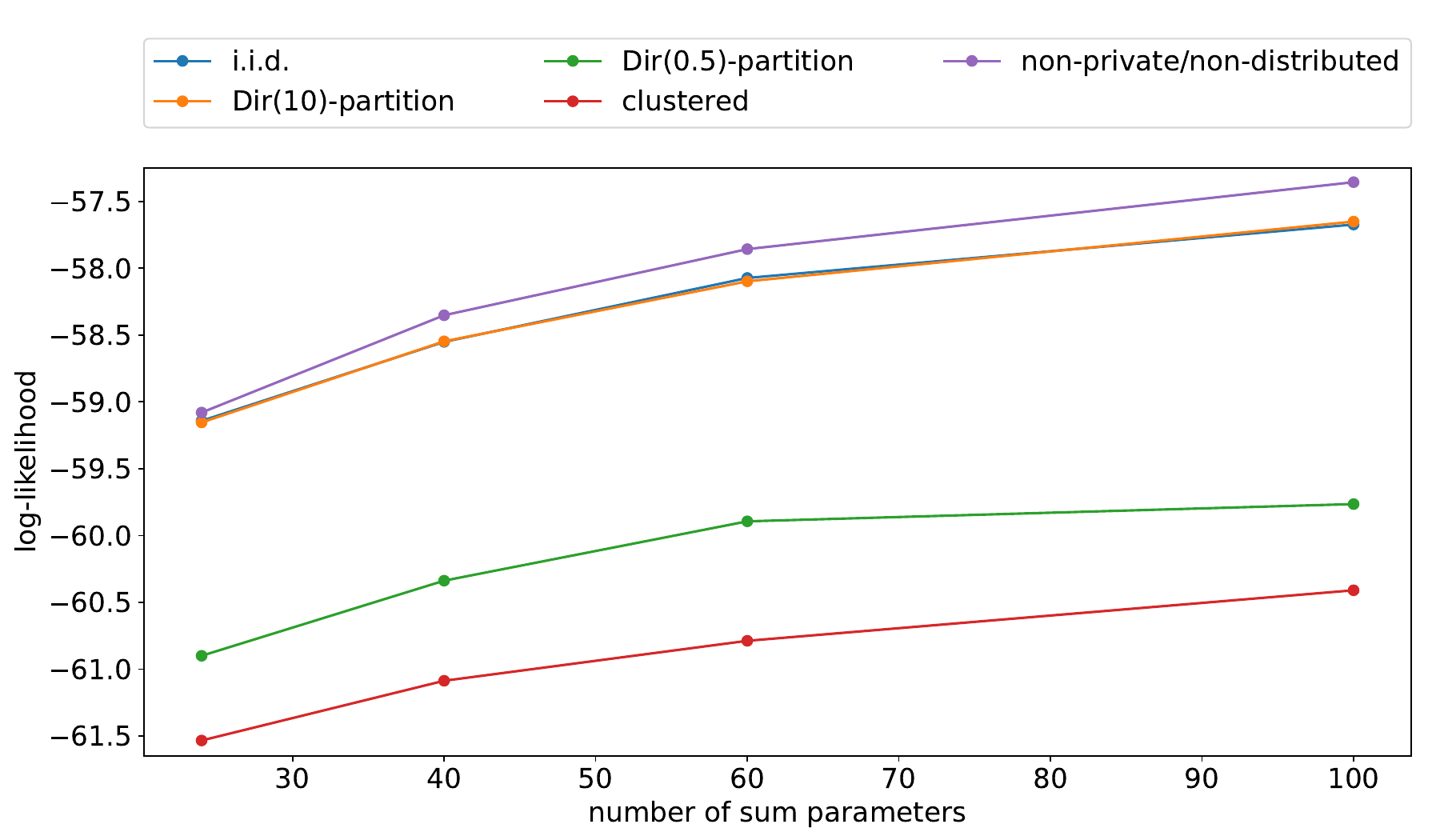} 
    \caption{Mean \textbf{log-likelihood} performance of 10 runs on the bnetflix dataset with 3 members and a varying number of sum weights on different dataset partitions. Here, the dataset partitions with a weak level of imbalance lie close to each other, while the log-likelihood decreases visibly for the other partitions. All experimental results can be found in the Appendix in Tables~\ref{tab:all-lls-iid}, \ref{tab:all-lls} and \ref{tab:all-lls-np}. }
    \label{fig:ll}
\end{figure}
\subsection{Log-likelihood Performance}
We note that before our contribution, no algorithm for a secure multi-party training scheme for SPNs had been proposed. Thus, a performance comparison in terms of log-likelihood with other methodologies is not easily possible or adequate. However, we provide experimental results for two baselines, namely a non-private, non-distributed version as well as a distributed, non-private version of our protocol. Here, the first baseline is a slight variation of the original RAT-SPN implementation by \cite{RAT-SPN}, as we implemented it on a forest of structures which each consist of less parameters than during their experiments. It yields results (see Table~\ref{tab:all-lls-iid}) which are close to the results provided for RAT-SPNs by \cite{RAT-SPN}. Thus, by comparing this baseline with our protocol, we provide an indirect comparison to current state-of-the-art SPN learners (with Table~$1$ by \cite{RAT-SPN}). The second baseline aggregates locally trained structures after one round at the main server. These baselines were evaluated for all four different dataset partitions. We specifically provide these two comparisons to show the influence of (i) distributing the data and (ii) including privacy-preserving primitives on the log-likelihood performance of the trained SPN forest. 

In Figure~\ref{fig:ll-np}, the results for the baselines and our protocol are visualized for the i.i.d partitioned data as well as the $Dir(10)$-partitioned data. Here, the distributed, non-private baseline shows results which are very similar to the results in our protocol. This holds for all datasets and all data partitions, as can be seen in Tables~\ref{tab:all-lls-iid}, \ref{tab:all-lls} and \ref{tab:all-lls-np}.  Small differences here are due to the random initialization of the RAT-SPN structures, which constitute the SPN forest in our protocol. We therefore conclude that adding the privacy-preserving primitives to our protocol does not influence the log-likelihood performance.
Also, in Figure~\ref{fig:ll-np} we see that the non-private, non-distributed baseline performs slightly better than our current protocol, which is mainly due to the distributed setting of our protocol.

The level of imbalance in a data partition, on the other hand, influences the log-likelihood performance of our protocol. Our protocol is robust against data partitions with a weak level of imbalance. This is shown in Figure~\ref{fig:ll} for the bnetflix dataset, were the log-likelihood performance measured for the $Dir(10)$-partition is very close to the i.i.d. partitioned data. Yet, for more heterogeneous data partitions, the performance decreases visibly, and it is left for further research to define up to which imbalance level the protocol's performance stays stable. 
Additionally, both Figure~\ref{fig:ll-np} and Figure~\ref{fig:ll} show an increase in performance when the number of sum weights in the overall forest is raised. This suggests that choosing larger structures for the forest is preferable.  
\begin{figure}[t]
    \centering
     
    \includegraphics[width=0.6\columnwidth]{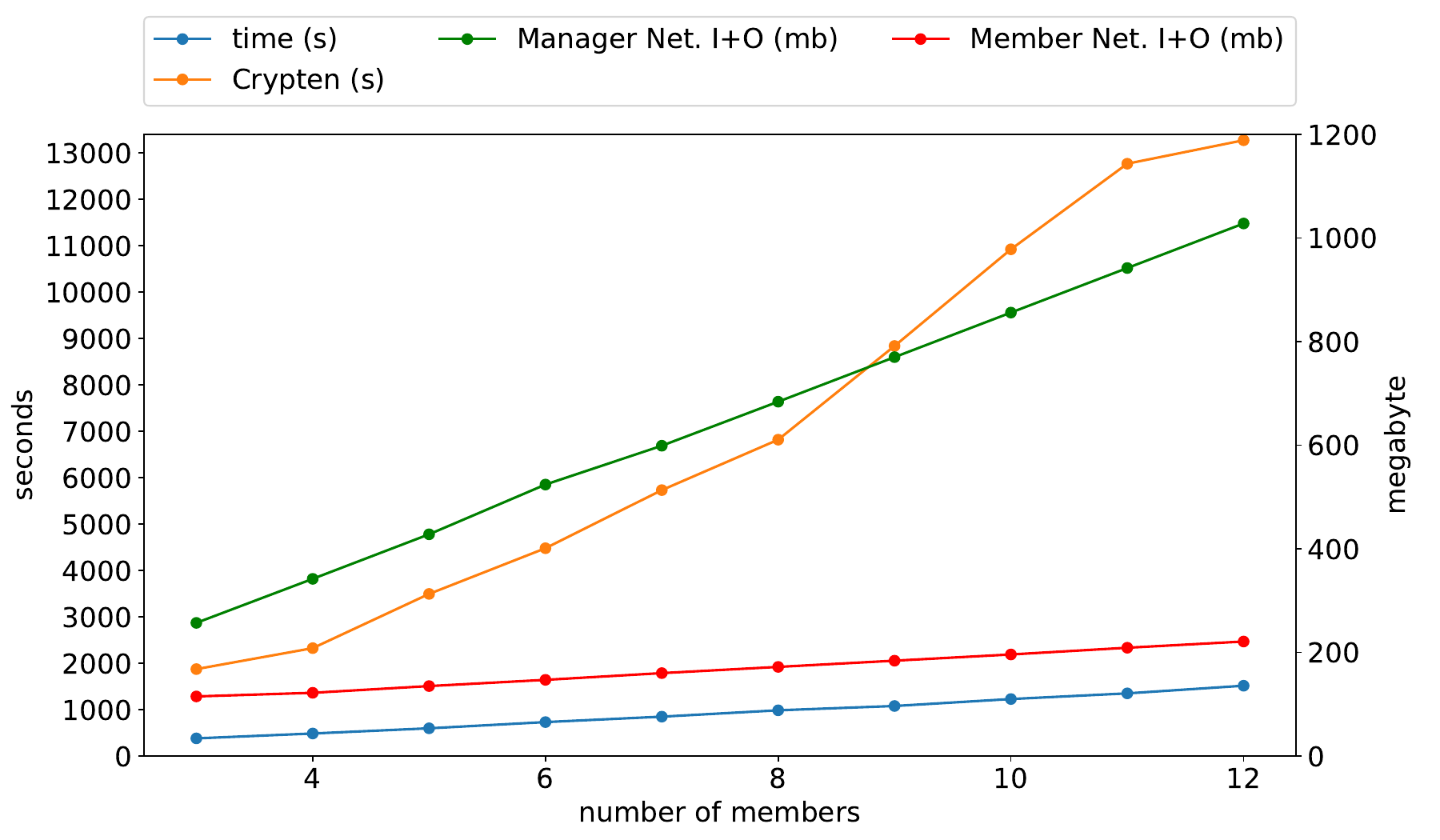}
   \caption{\textbf{Training} with network traffic for nltcs and a varying amount of members (408 trainable parameters, 30 local training epochs) with one manager and a latency of 10~ms. Training times for \texttt{CrypTen} (401 trainable parameters, 30 training epochs) are given as comparison (also see Table~\ref{tab:time} in the Appendix). Here, the \texttt{CrypTen} running times lie high above the results from our protocol. }
    \label{fig:time}
\end{figure}
\subsection{Time and Network Traffic}
For the scalability of the proposed approach, we evaluate run times and network traffic for 3 to 12 members. The results can be found in Figure~\ref{fig:time}. Training times and network traffic for the manager rise slowly when the number of members increases.
For 12 members, we report less than half an hour of training time on the nltcs dataset. Here, it is important to note that learning the model parameters is done only once during the lifetime of a model. Hence, running times in this order of magnitude for a given dataset are quite justifiable. 

To compare our approach to current multi-party approaches for DNNs, we trained a neural network via the \verb|CrypTen| software library \citep{crypten2020}. This library provides SMPC primitives over a PyTorch-like API and is the most widely used benchmark for SMPC. It guarantees security and privacy in the honest-but-curious setting and is, to the best of our knowledge, the only currently released software library able to handle secure and private neural network training over an arbitrary amount of parties (see \cite{crypten2020}, Table~3, p.~22)\footnote{Work developed for 2-4 parties include CryptGPU \citep{TanKTW21} or the Piranha platform \citep{Piranha}.}. We employ a multilayer perceptron-like architecture with 3 hidden layers and altogether 401 trainable parameters. Thus, the numbers of trainable parameters for \texttt{CrypTen} and our protocol are comparable. For the sake of simplicity, we train the neural network to reproduce the density estimates output by our SPN protocol. Nonetheless, we train this network for 30 epochs to keep the comparison to our protocol as fair as possible, and the training data is otherwise kept equal (see Table~\ref{tab:datastatistics} in the Appendix). In Figure~\ref{fig:time} we observe that this network requires significant training time, especially when compared to our SPN protocol. This suggests that our approach scales better with regard to running times. We also note that often considerably more than 30 epochs are necessary to bring a neural network model to convergence. This would increase the training runtime by approximately the same factor.
\begin{table}[t]
\centering
\footnotesize
\caption{10 \textbf{inference} calculations with network traffic for the nltcs dataset and a varying number of members with one manager and a latency of 10~ms. Network \,I+O\,(mb) is the network traffic of the manager and $\text{network \,I+O\,(mb)}^\ast$ of a single member. Times and network traffic for CryptoSPN are given as comparison.}
\begin{tabular}{c c c c c}
\toprule
\#\,parties&net.\,I+O\,(mb)&$\text{net.\,I+O\,(mb)}^\ast$&time\,(s)&CryptoSPN setup+\\
 & & & &online\,(s/mb)  \\ 
[0.5ex] 
\midrule
3 & 866 & 433 & 1306 & 63/440\\         %
5 & 1445 & 577 & 2238 & -\\          
10 & 2890 & 938 &  4601 & - \\      %

\bottomrule
\end{tabular}
\label{table:experimental13inference}
\end{table}
 \begin{table}[t]
\centering
\footnotesize
    \caption{Overview of the tasks executed during \textbf{training} on the nltcs dataset with 5 members, 3 structures (24 sum parameters, 384 leaves), 30 local training epochs, and a latency of 10~ms.}
	\begin{tabular}{c c c c }
		\toprule
	task		& \#\,executions			& time\,(s)	& time\,(\%) \\ \midrule
	truncation & 7745 & 239.31& 40.1\\
	multiplication& 3048& 99.29&16.6\\
	share\_values & 4675 & 91.87 &15.4\\
	addition & 4672 &81.25 & 13.6\\
	subtraction&2323&38.28 &6.4\\
	local\_structure\_training& 5 &30.91 & 5.2\\
	init\_network &5			& 8.37			& 1.4	 \\
	other & 234 &7.72& 1.3	\\ \bottomrule
	\end{tabular}
	\label{tab:percent}
\end{table} 

Inference times can be found for 3, 5 and 10 members on 10 data points in Table~\ref{table:experimental13inference}. We compare our results with the experimental results of CryptoSPN (see \cite{cryptospn}, Table 1) for the nltcs dataset and an SPN with 2 sum nodes and 640 leaf nodes. Further details of the comparison can be found in Appendix~\ref{apx:cryptoSPN}. The results show that the CryptoSPN protocol is very time efficient and costs less communication compared to our approach. Yet, in CryptoSPN the highest cost factor in the protocol is the setup complexity, which increases rapidly with the number of nodes in the underlying SPN (see \cite{cryptospn}, Table 1). This setup time is required each time when a new client wants to conduct inference for a datapoint, whereas in our protocol all clients taking part in training already hold all required information to directly infer datapoints. Furthermore, the CryptoSPN protocol cannot be used for multiple clients, whereas our protocol is constructed for an arbitrary number of clients and can additionally also handle inference queries from outside clients.
\begin{figure}[t]
    \centering
     
    \includegraphics[width=0.8\columnwidth]{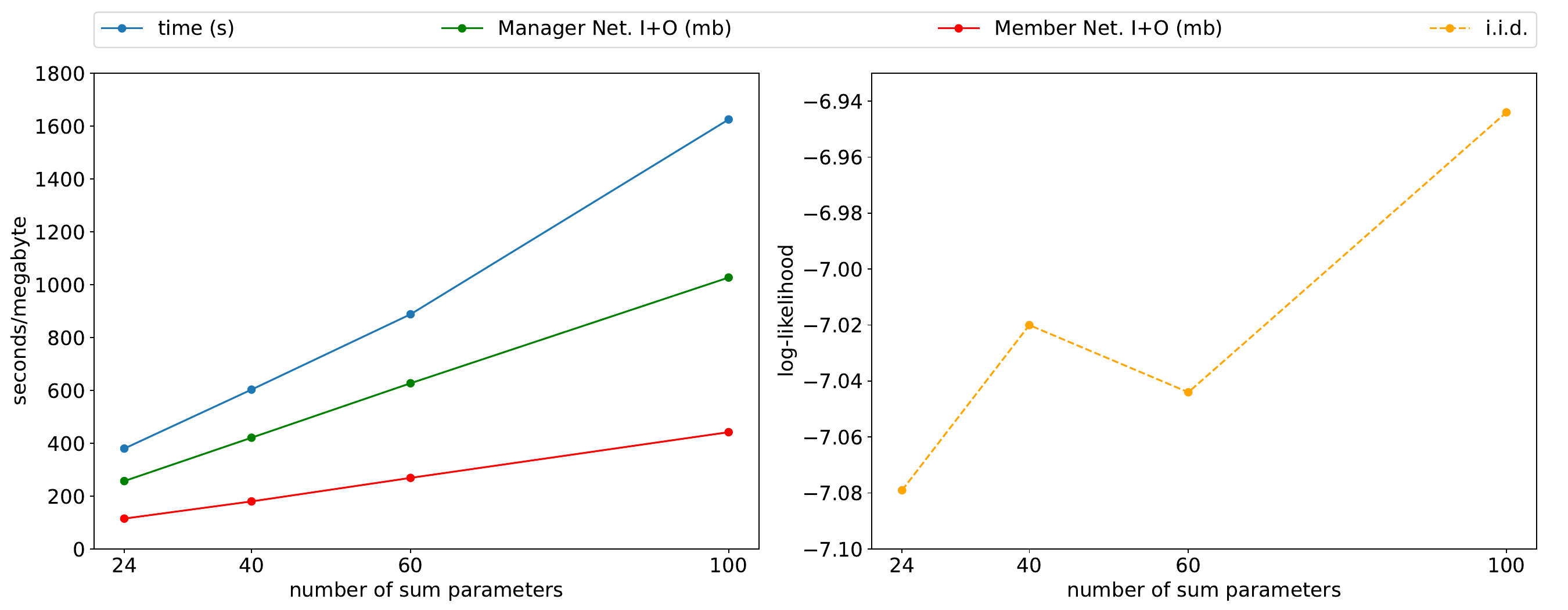}
   \caption{\textbf{Trend} for the log-likelihood performance (right) as well as the time and network traffic (left) when increasing the number of sum parameters in the SPN forest (also see Table~\ref{tab:tradeoff}). Experiment on the nltcs dataset with 3 members, one manager and a latency of 10~ms. We see that the log-likelihood improves for an increased number of sum parameters, but also that the time and network traffic increases.}
    \label{fig:trendsumnodes}
\end{figure}

Not only the number of members but also the size of the forest influences the time needed for training. In Figure~\ref{fig:trendsumnodes}, the impact of the size of the forest (with regard to the number of sum weights) on time and network traffic is depicted for three members. Both increase when the number of sum weights is raised. Furthermore, in Figure~\ref{fig:trendsumnodes}, we see that, when increasing the number of sum weights, we have a trade-off between the performance increase on the one hand and the time increase on the other hand. This is mostly due to the fact that for the learning of each sum parameter, a private division is needed, which influences the run time notably. This is underlined by the results shown in Table~\ref{tab:percent}, where the different tasks during the training with five members are depicted. Here, the truncation exercise, which is only needed for divisions, takes the longest time in the protocol. In future research, this trade-off could be reduced by parallelizing the divisions in the protocol.

\section{Discussion and Conclusion}	\label{sec:conc}
We see our work as a first contribution towards an efficient alternative to privacy-preserving DNNs. It tackles the challenges of today's privacy regulations and leads the way to further research on privacy-preserving SPNs. 
However, as this is the first attempt at addressing the task of a complete protocol for private structure and parameter learning for SPNs,  an experimental comparison with other methods is not directly possible. Our results suggest that private SPN learning can be done for realistically sized scenarios (for vectorial data, up to 135 features, 3 to 12 parties) end-to-end in under an hour. Compared to homomorphic encryption on DNNs, a recent approach required 30 hours of training on a much smaller dataset \citep{DNNTrainingHE}. For a result in the same range, but with larger data and based on secret sharing \citep{crypten2020}, we reported training times for the \verb|CrypTen| library in the previous section. Here, results turned out to be favorable for the approach presented here. 
Practical applicability is improved by the Docker implementation of our approach.

We also considered private inference, a protocol that has been solved previously using another approach (Yao's garbled circuits) by CryptoSPN \citep{cryptospn}. The latter approach is inherently limited to two parties (see also Table~\ref{table:experimental13inference}), while our approach enables multiple parties with heterogeneous data partitions to interact, which should further improve practical applicability.
%

As a next step, we would like to reduce the time needed for training and inference by parallelizing the divisions in our protocol. 
This would reduce the trade-off between log-likelihood performance and time/network traffic so that larger forests with a potentially better performance can be considered.
Parallelization can also be beneficial for inference, where this operation is applied to several data points at the same time. 
Additionally, other methods, such as for classification tasks, can be implemented into the protocol in the future.





\bibliography{refs}
\bibliographystyle{tmlr}
\newpage
\appendix
\section{SMPC: Addition and Multiplication} \label{app:SMPCadd+multi}
For additions and multiplications we apply existing basic protocols. 
Polynomial secret sharing is additive: Let $a,b \in \ZZ_p$ be two secrets, and $\{g(n)\}_{n=1}^N$ and $\{f(n)\}_{n=1}^N$ be their secret shares, respectively. Then $\{f(n) + g(n)\}_{n=1}^N$ are shares of $a+b$ (note that additions and multiplications are modulo $p$).

 Multiplying two shares intuitively could be performed similar to the above addition by letting each party $n$ locally multiply their shares $ f(n)g(n)$. However, if both $f$ and $g$ are of degree $t$, this results after multiplication in a polynomial of degree $2t$. Therefore, after some multiplications the degree of the secret polynomial is too high to perform further secret arithmetic operations over shares. Furthermore, the randomness of the polynomial is lost. To overcome this, we use the following approach by \cite{gennaro1998simplified}: Let again $f(n)$ and $g(n)$ denote the polynomial shares of $a,b$ of party $n$ with both polynomials of degree $t<N/2$. Then, there is a theorem proven by \cite{wigderson1988completeness} that for every $M$ we can compute numbers $(\lambda_1, \dots, \lambda_M)$ such that for each polynomial $f$ of degree at most $M-1$, it holds that $f(0)=\sum_{i=1}^{M} \lambda_i f(i)$. UUsing this theorem, a multiplication over shares can be performed in two steps: 
\begin{enumerate}
    \item Each party $n$ chooses a random polynomial $h_n(x)$ of degree $t$ such that $h_n(0) = f(n)g(n)$. Then $h_n$ is shared with all other parties by distributing $h_n(j)$ to party $j$ for $1\leq j \leq 2t+1$. 
    \item Each party $j$ then locally computes its share of $ab$ over the polynomial $H(x)=\sum_{n=1}^{2t+1}\lambda_n h_n(x)$ 
    by adding the previously received shares to $H(j) = \sum_{n=1}^{2t+1} \lambda_n h_n(j)$. 
       \end{enumerate}
Notice that $H$ is a polynomial of degree $t$ and $H(0)=\sum_{n=1}^{2t+1}\lambda_n h_n(0) =\sum_{n=1}^{2t+1}\lambda_n (f(n)g(n))=f(0)g(0)$.

\section{Additional Implementation Details}    \label{apx:impl}

In the implementation of our protocol, there is a task scheduling server called the \texttt{Manager}, and multiple task executing servers called \texttt{Member}s. They are all connected to each other via the WebSocket\footnote{\url{https://websockets.readthedocs.io/en/stable/}} Framework. All members and the manager have a unique \texttt{ID} in the network, which is used for secret sharing. Tasks to be done by the network are scheduled by the manager as \texttt{Exercise}s.

For our experiments, we used Docker at a computer with specification Ubuntu 22.04.1 LTS, 64~GB RAM and Ryzen Threadripper 1920X 12-Core Processor as CPU and an internal network latency of 10~ms at each all-to-all communication in the network. For secure arithmetic operations the parameter $truncate\_n$ for the Newton iterations and the truncation is set to  $\lceil \log(d) \rceil$, and the precision parameter $t$ to $24$. Our multiplicative factor $d$ to normalize real values is set to $10^7$, and as prime number we are using $p=2^{89}-1=618970019642690137449562111$. 

In the following experiments we focus on five datasets : nltcs, plants, baudio, bnetflix, and tretail \citep{datasets, tretail}. Statistics on these datsets are given in Table~\ref{tab:datastatistics}.
Furthermore, we look at homogeneous as well as heterogeneous distributions of these datasets across the clients. For the homogeneous setting, the data is independent and identically distributed (i.i.d.) in all members. Since this is seldom the case in real-world scenarios \citep{non-iid}, we also measure the performance for three heterogeneous data partitions. For this, we first use K-means to cluster the dataset with K equal to the number of members. Then, we distribute these clusters across the members simulating a weak, strong and very strong level of imbalance in the resulting data partitioning. Here, we apply a method commonly used to evaluate Federated Learning algorithms \citep{non-iid, Dir-partition}, which distributes a portion of the instances of cluster $k$ to party $n$ with a probability $p_k \sim Dir_n(\beta)$, where $Dir(\cdot)$ is the Dirichlet distribution. The concentration parameter $\beta$ influences the imbalance level, with larger values resulting in a more unbalanced data partition. Concretely, we choose $\beta=[0.5, 10]$ to simulate a weak ($Dir(10)$) and strong ($Dir(0.5)$) imbalance level and distribute a single cluster to each member to simulate the strongest level of imbalance. 

All experiments were run 10 times with a forest of 3-layered structures and a varying number of sum weights. Additionally, we evaluate for the smallest forest a variety of eight different local training criteria and three different weighting methods. 

 The first criteria are based on the number of local iterations. Here, the SPN is chosen which performs best on the validation set during training for 20, 30, 40, 100 and 300 local training iterations. Additionally, results are shown for different stopping criteria on the validation log-likelihood during training. In particular, local training stops if the absolute value of the difference between the current log-likelihood on the validation set and the best log-likelihood on the validation set so far is less than $\mu$ for the current iteration and the following 10 iterations. For $\mu$, the values 0.001, 0.0005 and 0.0001 were analyzed.

 \begin{table}[t]
    	\centering
    		\caption{Dataset statistics on five datasets \cite{datasets, tretail} used during the conducted experiments.}
	\begin{tabular}{ccccc}
	\toprule

	dataset		& \#vars			& \#train + \#validation &  \#global test & density \\ \midrule
		nltcs			& 8			& 19253	    & 2093      & 0.332	\\ 
		plants			& 69	 	& 20894 	& 2321      & 0.180	\\ 
		baudio			& 100 		&18000      & 2000      &0.199 \\
		bnetflix		& 100		&18000 	    & 2000      &0.541	\\ 
		tretail			& 135	    & 26449 	& 2938	    &0.024	\\ \bottomrule
	\end{tabular}

	\label{tab:datastatistics}
\end{table}

 As for the weighting of the SPNs in the forest, we evaluated uniform weighting as well as three different methods based on log-likelihood performance. The ranking method is already explained in detail in Section~\ref{sec:forestweight}. Furthermore, we conducted experiments on a log-likelihood weighting method and an enforced log-likelihood weighting method. For the first method, the local structure weights were calculated by
\begin{equation}\label{eq:llmethod}
\centering
    s_k^n =  \frac{1}{\lvert \ell \ell_n[k]\rvert},
\end{equation}
 with $\lvert\ell \ell_n[k]\rvert$ as the log-likelihood of structure $\mathcal{S}^n_k$ on party $n$'s validation data. For the second method, Equation~\ref{eq:llmethod} was modified to 
\begin{equation}
    s_k^n =  \frac{1}{\lvert\ell \ell_n[k]\rvert^2}
\end{equation}
in order to enlarge the difference between the structure weights. For both methods, the secret sharing of the parameters is realized as for the weighting method explained in Section~\ref{sec:forestweight}.

\begin{figure}[t]
    \centering
     
    \includegraphics[width=\columnwidth]{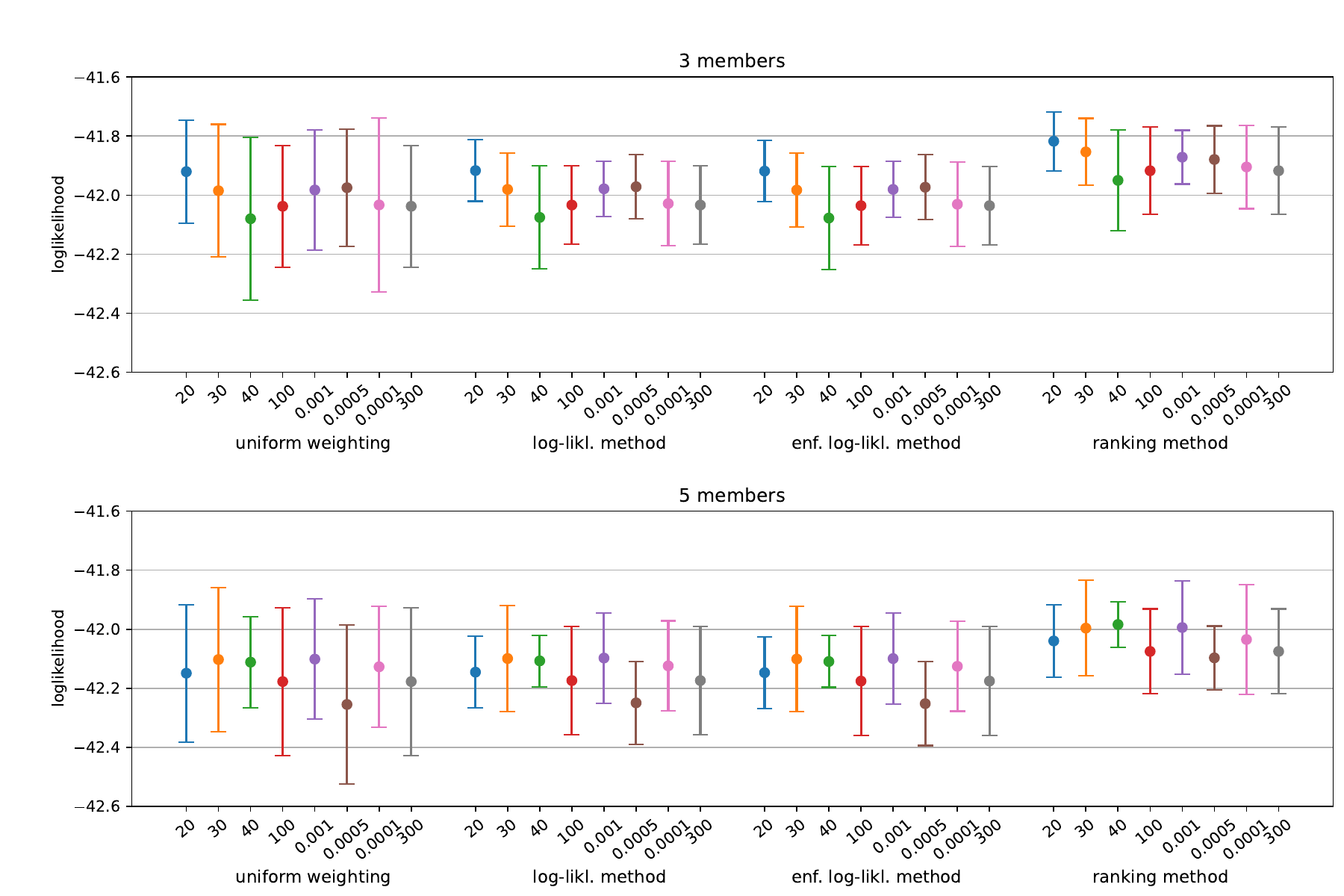}
   \caption{Mean log-likelihood and standard deviation for the different weighting methods each with varying training criteria (20, 30, 40, 100, 300 local training iterations as well as 0.001, 0.0005 and 0.0001 as stopping criteria), 3 members (mean of 20 runs) and 5 members (mean of 10 runs) on the i.i.d.\ partitioned baudio dataset.}
    \label{fig:baudiotc}
\end{figure}

As visualized for the baudio dataset and the i.i.d. partitioned data in Figure~\ref{fig:baudiotc}, uniform weighting gives, as expected, the results with the highest variance. Overall, the difference in mean performance for the local training methods is very small. However, in terms of weighting methods, the ranking method clearly shows empirically the best performance (both in variance and the mean) and was therefore used as standard weighting method in the experiments conducted in Section~\ref{sec:exp}. 
 \section{Comparison to CryptoSPN} \label{apx:cryptoSPN}
To compare the CryptoSPN protocol by \cite{cryptospn} with our approach we firstly reran the CryptoSPN protocol on our local machine with the same input parameters as for our protocol. But, as the results differed largely from the results given in Table 1 of the CryptoSPN paper we decided to take the paper as reference. We believe that the deviation was mainly due to a different network setting as well as different restrictions for communication. Furthermore, experiments for CryptoSPN were run on two machines with 128 GB RAM and Intel Core i9-7960X 16-core Processor as CPU \citep{cryptospn} compared to one machine with 64 GB RAM and Ryzen Threadripper 1920X 12-Core Processor as CPU on which we reran the CryptoSPN experiments. To set our protocol in relation with CryptoSPN we have the following differences in parameter choice. Our protocol's security parameter is based on the prime number used, so $89$-bit security, whereas CryptoSPN uses a security parameter of $128$-bits. Additionally, our precision is $t -5 -\log(N+1)$ (see \cite{algesheimer2002efficient}) and therefore $18$-bit compared to $32$-bit for CryptoSPN.

\begin{table*}[h!]
    \small
     \centering
     \caption{Mean log-likelihood of 10 runs for 3 and 5 members on 5 datasets with a varying structural forest. Results are shown for our two baselines (non-private, non-distributed baseline and non-private, distributed baseline) as well as the results of the private, i.i.d. setting. Best results for each setting with regard to the number of sum parameters are in bold. We can see that in general more sum parameters improve the log-likelihood. }
	\begin{tabularx}{0.9\textwidth}{c|ccccc}
	
	\toprule
		dataset		& \#parties&\#sum params & i.i.d. & non-private & non-private/\\
		&&&&&non-distributed \\ \midrule
	    \multirow{2}{*}{nltcs}		
	&\multirow{4}{*}{3}& 24 	&-7.079 ($\pm$ 0.215)&-7.240 ($\pm$ 0.348)& -6.164\\ 
	&&40&-7.020 ($\pm$ 0.223)&-7.120 ($\pm$ 0.145) &-6.085\\
&& 60&

-7.044 ($\pm$ 0.111)&-7.092 ($\pm$ 0.162)&-6.052 \\
&& 100&

\textbf{-6.944} ($\pm$ 0.188)&\textbf{-6.975} ($\pm$ 0.140)&\textbf{-6.029}\\\cline{2-6}
	
		&\multirow{4}{*}{5}& 24	&-7.976 ($\pm$ 0.286) &-8.036 ($\pm$ 0.232)&\\
		&& 40&-8.042 ($\pm$ 0.162)&-8.015 ($\pm$ 0.158)& \\
&& 60&
-8.018 ($\pm$ 0.158) &-8.070 ($\pm$ 0.155)
 \\
&& 100&\textbf{-7.968 }($\pm$ 0.199)&\textbf{-7.917} ($\pm$ 0.109)&\\\midrule

	\multirow{2}{*}{plants}		
		&\multirow{4}{*}{3} 
		&24&-17.868 ($\pm$ 0.164)&-17.859 ($\pm$ 0.213)&-17.710\\
		&&40&-16.754 ($\pm$ 0.194)&-16.895 ($\pm$ 0.088)&-16.072\\
	&&60&-16.092 ($\pm$ 0.138)&-16.052 ($\pm$ 0.103)&-15.226\\
	&&100&\textbf{-15.489 }($\pm$ 0.108)&\textbf{-15.522} ($\pm$ 0.086)&\textbf{-14.404}\\ \cline{2-6}
		&\multirow{4}{*}{5}&24 	&-18.048 ($\pm$ 0.275)&-18.162 ($\pm$ 0.233)&\\
		&&40&-17.092 ($\pm$ 0.158)&-17.116 ($\pm$ 0.172)\\
	&&60&-16.345 ($\pm$ 0.168)&-16.414 ($\pm$ 0.151)\\
	&&100&\textbf{-15.781 }($\pm$ 0.106)&\textbf{-15.855 }($\pm$ 0.080)\\ \midrule
		
		\multirow{2}{*}{baudio}		
		&\multirow{4}{*}{3}  &24	&-41.817 ($\pm$ 0.100)&-41.825 ($\pm$ 0.064)&-41.832\\ 
	&& 40&
	-41.312 ($\pm$ 0.090)&-41.302 ($\pm$ 0.093)&-41.083 \\
&& 60&
-40.969 ($\pm$ 0.045)&-40.961 ($\pm$ 0.036)&-40.672 \\
&& 100&
\textbf{-40.688} ($\pm$ 0.043) &\textbf{-40.688} ($\pm$ 0.048)&\textbf{-40.267}\\

 \cline{2-6}
		&\multirow{4}{*}{5} &24	&-41.984 ($\pm$ 0.076)&-42.081 ($\pm$ 0.116)&\\
		&&40&-41.520 ($\pm$ 0.091)&-41.529 ($\pm$ 0.095) \\
&& 60&
-41.204 ($\pm$ 0.058) &-41.248 ($\pm$ 0.064)\\
	&&100&\textbf{-41.117} ($\pm$ 0.049)&\textbf{ -41.134 }($\pm$ 0.043)\\ \midrule
		
		\multirow{2}{*}{bnetflix}		
		&\multirow{4}{*}{3}&24 	&-59.142 ($\pm$ 0.142)& -59.150 ($\pm$ 0.114)&-59.080\\ 
		&& 40&-58.551 ($\pm$ 0.123) &-58.520 ($\pm$ 0.102)&-58.352\\
&& 60&

-58.073 ($\pm$ 0.090)&-58.060 ($\pm$ 0.076)&-57.857 \\
&& 100&

\textbf{-57.673 }($\pm$ 0.076) &\textbf{-57.674 }($\pm$ 0.061)&\textbf{-57.356}
\\ \cline{2-6}
		&\multirow{4}{*}{5}&24   	&-59.243 ($\pm$ 0.164)&-59.322 ($\pm$ 0.145)\\
	&& 40&-58.679 ($\pm$ 0.069)&-58.733 ($\pm$ 0.097)\\
&& 60&

-58.217 ($\pm$ 0.079)&-58.247 ($\pm$ 0.077)\\
	&&100&\textbf{-57.904} ($\pm$ 0.058)&\textbf{-57.851} ($\pm$ 0.034)\\\midrule
		
		\multirow{2}{*}{tretail}		
		&\multirow{4}{*}{3} & 24	&-11.055 ($\pm$ 0.022)&-11.072 ($\pm$ 0.019)&-10.931\\ 
		&&40&-11.054 ($\pm$ 0.020)&-11.050 ($\pm$ 0.018)&-10.875\\
	&&60&-11.031 ($\pm$ 0.013)& \textbf{-11.028 }($\pm$ 0.013)&-10.868\\
	&&100&\textbf{-11.018} ($\pm$ 0.007)&-11.029 ($\pm$ 0.011)&\textbf{-10.862}\\ \cline{2-6}
		&\multirow{4}{*}{5}&24 	&-11.071 ($\pm$ 0.020)&-11.092 ($\pm$ 0.031)\\ 
		&&40&-11.066 ($\pm$ 0.012)&-11.067 ($\pm$ 0.020)\\
	&&60&-11.060 ($\pm$ 0.012)&-11.063 ($\pm$ 0.015)\\
	&&100&\textbf{-11.050} ($\pm$ 0.014)&\textbf{-11.056} ($\pm$ 0.016)\\
	\bottomrule
	\end{tabularx}
	\label{tab:all-lls-iid}
\end{table*}

\begin{table*}[h]
     \small
     \centering
     \caption{Mean log-likelihood of 10 private runs for 3 and 5 members on 5 differently partitioned datasets with a varying structural forest.  Best results for each setting with regard to the number of sum parameters are in bold. We can see that for most datasets more sum parameters improve the log-likelihood.  }
	\begin{tabularx}{0.9\textwidth}{c|ccccc}
	
	\toprule
		dataset		& \#parties&\#sum params  & Dir(10)& Dir(0.5) &clustered  \\ \midrule
	    \multirow{2}{*}{nltcs}		
	&\multirow{4}{*}{3}& 24 	&-7.522 ($\pm$ 0.244)&-7.627 ($\pm$ 0.202)&\textbf{-9.113} ($\pm$ 0.134)\\ 
	&&40&\textbf{-7.146} ($\pm$ 0.129)&\textbf{-7.549} ($\pm$ 0.244)&-9.187 ($\pm$ 0.048) \\
&& 60&

-7.211 ($\pm$ 0.138)&-7.666 ($\pm$ 0.109)&-9.216 ($\pm$ 0.080)\\
&& 100
&-7.331 ($\pm$ 0.182)&-7.606 ($\pm$ 0.126)&-9.146 ($\pm$ 0.079) \\\cline{2-6}
	
		&\multirow{4}{*}{5}& 24	 &-8.073 ($\pm$ 0.217)&-8.876 ($\pm$ 0.143)&\textbf{-9.391 }($\pm$ 0.083)\\
		&& 40&\textbf{-8.019} ($\pm$ 0.178)&\textbf{-8.706} ($\pm$ 0.121)&-9.395 ($\pm$ 0.041) \\
&& 60&
-8.059 ($\pm$ 0.193)&-8.719 ($\pm$ 0.156)&-9.405 ($\pm$ 0.047)
 \\
&& 100&-8.125 ($\pm$ 0.209)&-8.845 ($\pm$ 0.078)&-9.428 ($\pm$ 0.023)\\\midrule

	\multirow{2}{*}{plants}		
		&\multirow{4}{*}{3} &24  	&-18.009 ($\pm$ 0.284)&-21.254 ($\pm$ 0.557)&-29.382 ($\pm$ 0.627)\\
		&&40&-16.997 ($\pm$ 0.209)&-20.107 ($\pm$ 0.244)&-28.706 ($\pm$ 0.256)\\
	&&60&-16.213 ($\pm$ 0.138)&-19.482 ($\pm$ 0.302)&-28.186 ($\pm$ 0.297)\\
	&&100&\textbf{-15.609} ($\pm$ 0.118)&\textbf{-19.003} ($\pm$ 0.252)&\textbf{-27.841 }($\pm$ 0.264)\\ \cline{2-6}
		&\multirow{4}{*}{5}&24 &-18.622 ($\pm$ 0.176)&-22.093 ($\pm$ 0.232)&-31.820 ($\pm$ 0.584)\\
		&&40&-17.560 ($\pm$ 0.182)&-20.577 ($\pm$ 0.266)&-31.171 ($\pm$ 0.331)\\
	&&60&-16.806 ($\pm$ 0.175)&-19.956 ($\pm$ 0.300)&-31.373 ($\pm$ 0.357)\\
	&&100&\textbf{-16.408} ($\pm$ 0.131)&\textbf{-19.840 }($\pm$ 0.194)&\textbf{-30.920} ($\pm$ 0.318)\\ \midrule
		
		\multirow{2}{*}{baudio}		
		&\multirow{4}{*}{3}  &24&-41.901 ($\pm$ 0.106)&-43.601 ($\pm$ 0.242)&-47.013 ($\pm$ 0.355)\\ 
	&& 40&-41.378 ($\pm$ 0.099)&-43.314 ($\pm$ 0.199)&-46.573 ($\pm$ 0.271) \\
&& 60&-41.037 ($\pm$ 0.055)&-43.088 ($\pm$ 0.256)&-46.329 ($\pm$ 0.140) \\
&& 100 &\textbf{-40.749} ($\pm$ 0.045)&\textbf{-42.726 }($\pm$ 0.112)&\textbf{-46.011 }($\pm$ 0.183)\\

 \cline{2-6}
		&\multirow{4}{*}{5} &24	&-42.159 ($\pm$ 0.135)&-43.110 ($\pm$ 0.218)&-48.366 ($\pm$ 0.232)\\
		&&40&-41.629 ($\pm$ 0.097)&-42.535 ($\pm$ 0.132)&-47.960 ($\pm$ 0.270) \\
&& 60&-41.281 ($\pm$ 0.092)&-42.220 ($\pm$ 0.134)&-47.824 ($\pm$ 0.200)\\
	&&100&\textbf{-41.115} ($\pm$ 0.035)&\textbf{-42.148} ($\pm$ 0.128)&\textbf{-47.583 }($\pm$ 0.104)\\ \midrule
		
		\multirow{2}{*}{bnetflix}		
		&\multirow{4}{*}{3}&24 &-59.154 ($\pm$ 0.139)&-60.900 ($\pm$ 0.184)&-61.534 ($\pm$ 0.186) \\ 
		&& 40&-58.547 ($\pm$ 0.056)&-60.338 ($\pm$ 0.225)&-61.087 ($\pm$ 0.174)\\
&& 60&-58.098 ($\pm$ 0.066)&-59.894 ($\pm$ 0.116)&-60.788 ($\pm$ 0.076) \\
&& 100&\textbf{-57.651} ($\pm$ 0.024)&\textbf{-59.765} ($\pm$ 0.095)&-\textbf{60.410 }($\pm$ 0.077)
\\ \cline{2-6}
		&\multirow{4}{*}{5}&24   	&-59.392 ($\pm$ 0.093)&-60.677 ($\pm$ 0.190)&-62.237 ($\pm$ 0.136)\\
	&& 40&-58.851 ($\pm$ 0.120)&-60.106 ($\pm$ 0.141) &-61.864 ($\pm$ 0.112) \\
&& 60&-58.403 ($\pm$ 0.068)&-59.678 ($\pm$ 0.136) &-61.573 ($\pm$ 0.099)\\
	&&100&\textbf{-58.028} ($\pm$ 0.082)&\textbf{-59.765} ($\pm$ 0.095)&\textbf{-61.286} ($\pm$ 0.063)\\\midrule
		
		\multirow{2}{*}{tretail}		
		&\multirow{4}{*}{3} & 24	&-11.085 ($\pm$ 0.025)&-11.258 ($\pm$ 0.025)&-11.362 ($\pm$ 0.010)\\ 
		&&40&-11.064 ($\pm$ 0.018)&-11.248 ($\pm$ 0.012)&-11.356 ($\pm$ 0.011)\\
	&&60&-11.055 ($\pm$ 0.017)&-11.237 ($\pm$ 0.019)&-11.353 ($\pm$ 0.014)\\
	&&100&\textbf{-11.041 }($\pm$ 0.015)&\textbf{-11.214} ($\pm$ 0.014)&\textbf{-11.333 }($\pm$ 0.004)\\ \cline{2-6}
		&\multirow{4}{*}{5}&24 	&-11.143 ($\pm$ 0.014)&-11.333 ($\pm$ 0.010)&\textbf{-11.363} ($\pm$ 0.016) \\ 
		&&40&-11.116 ($\pm$ 0.017)&-11.285 ($\pm$ 0.010)&-11.401 ($\pm$ 0.007)\\
	&&60&-11.095 ($\pm$ 0.018)&
\textbf{-11.268} ($\pm$ 0.010)&-11.369 ($\pm$ 0.049)\\
	&&100&\textbf{-11.090} ($\pm$ 0.014)&-11.329 ($\pm$ 0.021)&-11.376 ($\pm$ 0.016)\\ \bottomrule
	\end{tabularx}
	
	\label{tab:all-lls}
\end{table*}

\begin{table*}[h!]
     \small
     \centering
     	\caption{ Mean log-likelihood of 10 non-private runs for 3 and 5 members on 5 differently partitioned datasets with a varying structural forest.  Best results for each setting with regard to the number of sum parameters are in bold. We can see that for most datasets more sum parameters improve the log-likelihood. }
	\begin{tabularx}{0.9\textwidth}{c|ccccc}
	
	\toprule
		dataset		& \#parties&\#sum params  & Dir(10)& Dir(0.5) &clustered  \\ \midrule
	    \multirow{2}{*}{nltcs}		
	&\multirow{4}{*}{3}& 24 	&\textbf{-7.254} ($\pm$ 0.237)&-7.572 ($\pm$ 0.271)&-9.156 ($\pm$ 0.091)\\ 
	&&40&-7.348 ($\pm$ 0.322)&\textbf{-7.517} ($\pm$ 0.196)&-9.193 ($\pm$ 0.074)\\
&& 60&

\textbf{-7.254} ($\pm$ 0.159)&-7.661 ($\pm$ 0.136)&\textbf{-9.144} ($\pm$ 0.060)\\
&& 100
&-7.309 ($\pm$ 0.147)&-7.593 ($\pm$ 0.145)&-9.178 ($\pm$ 0.045) \\\cline{2-6}
	
		&\multirow{4}{*}{5}& 24	 &-8.312 ($\pm$ 0.168)&-8.810 ($\pm$ 0.070)&-9.451 ($\pm$ 0.036)\\
		&& 40&-8.112 ($\pm$ 0.223)&-8.826 ($\pm$ 0.146)&-9.435 ($\pm$ 0.034) \\
&& 60&
\textbf{-8.092} ($\pm$ 0.195)&\textbf{-8.778} ($\pm$ 0.070)&\textbf{-9.391} ($\pm$ 0.033)
 \\
&& 100&\textbf{-8.092} ($\pm$ 0.186)&-8.832 ($\pm$ 0.099)&-9.412 ($\pm$ 0.029)\\\midrule

	\multirow{2}{*}{plants}		
		&\multirow{4}{*}{3} &24  	&-17.994 ($\pm$ 0.213)&-21.278 ($\pm$ 0.655)&-29.567 ($\pm$ 0.484)\\
		&&40& -16.948 ($\pm$ 0.071)&-20.227 ($\pm$ 0.353)&-28.777 ($\pm$ 0.479)\\
	&&60&-16.214 ($\pm$ 0.143)&-19.504 ($\pm$ 0.275)&-28.036 ($\pm$ 0.316)\\
	&&100&\textbf{-15.587} ($\pm$ 0.082)&\textbf{-18.991} ($\pm$ 0.219)&\textbf{-27.834} ($\pm$ 0.195)\\ \cline{2-6}
		&\multirow{4}{*}{5}&24 &-18.412 ($\pm$ 0.217)&-22.043 ($\pm$ 0.719)&-31.846 ($\pm$ 0.472)\\
		&&40&-17.444 ($\pm$ 0.179)&-20.664 ($\pm$ 0.395)&-31.454 ($\pm$ 0.406)\\
	&&60&-16.923 ($\pm$ 0.152)&-20.270 ($\pm$ 0.273)&-31.100 ($\pm$ 0.351)\\
	&&100&\textbf{-16.369} ($\pm$ 0.153)&\textbf{-19.920} ($\pm$ 0.242)&\textbf{-30.712} ($\pm$ 0.301)\\ \midrule
		
		\multirow{2}{*}{baudio}		
		&\multirow{4}{*}{3}  &24&-41.927 ($\pm$ 0.124)&-44.021 ($\pm$ 0.293)& -47.036 ($\pm$ 0.282)\\ 
	&& 40&-41.378 ($\pm$ 0.094)&-43.305 ($\pm$ 0.279)& -46.422 ($\pm$ 0.248)\\
&& 60&-41.020 ($\pm$ 0.065)&-42.926 ($\pm$ 0.240)& -46.322 ($\pm$ 0.178) \\
&& 100 &\textbf{-40.732 }($\pm$ 0.051)&\textbf{-42.814} ($\pm$ 0.163)&\textbf{-46.06}2 ($\pm$ 0.142)\\

 \cline{2-6}
		&\multirow{4}{*}{5} &24	&-43.108 ($\pm$ 0.136)&-43.110 ($\pm$ 0.218)&-48.325 ($\pm$ 0.402)\\
		&&40&-41.608 ($\pm$ 0.089)&-42.597 ($\pm$ 0.132)&-48.138 ($\pm$ 0.166) \\
&& 60&-41.293 ($\pm$ 0.065)&-42.286 ($\pm$ 0.094)& -47.730 ($\pm$ 0.191)\\
	&&100&\textbf{-41.105} ($\pm$ 0.036)&\textbf{-42.214} ($\pm$ 0.050)& \textbf{-47.658} ($\pm$ 0.085)\\ \midrule
		
		\multirow{2}{*}{bnetflix}		
		&\multirow{4}{*}{3}&24 &-59.207 ($\pm$ 0.117)&-60.901 ($\pm$ 0.132)&-61.645 ($\pm$ 0.119)\\ 
		&& 40&-58.517 ($\pm$ 0.104)&-60.354 ($\pm$ 0.109)&-61.087 ($\pm$ 0.120)\\
&& 60&-58.080 ($\pm$ 0.061)&-59.994 ($\pm$ 0.184)& -60.709 ($\pm$ 0.091) \\
&& 100&\textbf{-57.605} ($\pm$ 0.036)&\textbf{-59.742} ($\pm$ 0.108)&\textbf{-60.427} ($\pm$ 0.099)
\\ \cline{2-6}
		&\multirow{4}{*}{5}&24   	&-59.475 ($\pm$ 0.132)&-60.674 ($\pm$ 0.186)& -62.240 ($\pm$ 0.140)\\
	&& 40&-58.827 ($\pm$ 0.083)&-60.063 ($\pm$ 0.139)&-61.891 ($\pm$ 0.087) \\
&& 60&-58.401 ($\pm$ 0.116)&-59.697 ($\pm$ 0.138) &-61.569 ($\pm$ 0.053)\\
	&&100&\textbf{-58.035} ($\pm$ 0.064)&\textbf{-59.313} ($\pm$ 0.114)& \textbf{-61.282} ($\pm$ 0.072)\\\midrule
		
		\multirow{2}{*}{tretail}		
		&\multirow{4}{*}{3} & 24	&-11.083 ($\pm$ 0.015)&-11.272 ($\pm$ 0.020)&-11.365 ($\pm$ 0.018)\\ 
		&&40&-11.067 ($\pm$ 0.030)&-11.247 ($\pm$ 0.023)&-11.346 ($\pm$ 0.014)\\
	&&60&-11.054 ($\pm$ 0.012)&-11.231 ($\pm$ 0.017)& -11.349 ($\pm$ 0.013)\\
	&&100&\textbf{-11.040} ($\pm$ 0.016)&\textbf{-11.213} ($\pm$ 0.016)&\textbf{-11.333} ($\pm$ 0.011)\\ \cline{2-6}
		&\multirow{4}{*}{5}&24 	&-11.148 ($\pm$ 0.025)&-11.299 ($\pm$ 0.013)&-11.368 ($\pm$ 0.016) \\ 
		&&40&-11.119 ($\pm$ 0.017)&-11.279 ($\pm$ 0.011)&-11.351 ($\pm$ 0.008)\\
	&&60&-11.102 ($\pm$ 0.019)&
-11.267 ($\pm$ 0.006)&-11.350 ($\pm$ 0.007)\\
	&&100&\textbf{-11.091} ($\pm$ 0.009)&\textbf{-11.256} ($\pm$ 0.006)&\textbf{-11.344} ($\pm$ 0.005)\\ \bottomrule
	\end{tabularx}

	\label{tab:all-lls-np}
\end{table*}

\begin{table}[h]
     
     \centering
     \caption{\textbf{Training} with network traffic for nltcs, 3 members and a varying number of sum weights with one manager and a latency of 10~ms. Network I+O\,(mb) is the network traffic of the manager and $\text{Network I+O\,(mb)}^\ast$ of a single member.}
	\begin{tabular}{c c c c }
	
	\toprule
		\#\,sum weights & net. I+O\,(mb) & $\text{net. I+O\,(mb)}^\ast$ &  time\,(s) \\ [0.5ex] \midrule
	   		
	 24& 257 & 115 &   380 \\ 
	40 & 421 &180&603   \\
 60&627&269&888

 \\
100&1027 & 442 & 1625

 \\ \bottomrule 
\end{tabular}

\label{tab:tradeoff}
\end{table}

\begin{table}[h]
\centering
\caption{\textbf{Training} with network traffic for nltcs and a varying amount of members (408 trainable parameters, 30 local training epochs) with one manager and a latency of 10~ms. Network I+O\,(mb) is the network traffic of the manager and $\text{network I+O\,(mb)}^\ast$ of a single member. Training times for \texttt{CrypTen}(401 trainable parameters, 30 training epochs) are given as comparison.}
\begin{tabular}{c c c c c}
\toprule
\#\,parties & net. I+O\,(mb) & $\text{net. I+O\,(mb)}^\ast$ &  time\,(s) & \verb|CrypTen|  \\ [0.5ex] 
\midrule
 3 &   257 & 115 &   380 & 1873 \\
 4 &   342 & 122 &   485 & 2325 \\           %
 5 &   428 & 135 &   597 & 3494 \\          %
 6 &   514 & 147 &   731 & 4479 \\          %
 7 &   599 & 160 &   848 & 5732 \\           %
 8 &   684 & 172 &   985 & 6819 \\          %
 9 &   770 & 184 &   1078 &  8839\\          %
10 & 856 & 196 &   1228 & 10923 \\      %
11 & 941 & 209 &  1350 & 12768\\       %
12 & 1028 & 221 & 1515 & 13274 \\       %
 [1ex]
\bottomrule
\end{tabular}

\label{tab:time}
\end{table}

\end{document}